\documentclass[12pt]{article}
\usepackage{graphicx}
\begin{document}
\centerline{\Large \bf Asymmetry in hierarchy model of Bonabeau et al} 

\medskip
D. Stauffer$^1$ and J.S. S\'a Martins$^2$
 
\medskip
Laboratoire PMMH, Ecole Sup\'erieure de Physique et Chimie
Industrielle, 10 rue Vauquelin, F-75231 Paris, Euroland
 
\medskip
 
$^1$ Visiting from Institute for Theoretical Physics, Cologne
University, D-50923 K\"oln, Euroland;
stauffer@thp.uni-koeln.de
 
$^2$ Visiting from Instituto de F\'{\i}sica, Universidade
Federal Fluminense; Av. Litor\^{a}nea s/n, Boa Viagem,
Niter\'{o}i 24210-340, RJ, Brazil; jssm@if.uff.br

\medskip

Abstract: The 1995 model of Bonabeau et al. explains the emergence of social 
hierarchies through randomness, but gives as many leaders as followers.
A simple modification allows a more realistic asymmetry with much less leaders.

\medskip

Ref. \cite{bonabeau} introduced an explanation, henceforth called the Bonabeau
model, to explain the emergence of social hierarchies through randomness: If
two people compete for the same living space, randomly one of them wins and
one of them loses. The more victories (losses) a person had in the recent
past, the higher (lower) is the probability to win the duel. With suitable
modifications \cite{sousa,stauffer} an equilibrium phase transition was
found at a concentration near 0.32 between egalitarian society at low 
population density and hierarchical society at high population density.
Schweitzer \cite{schweitzer} has criticized that in this model there are as many
leaders as there are followers, in contrast to reality. This symmetry is avoided
in the present modification.

\begin{figure}[hbt]
\begin{center}
\includegraphics[angle=-90,scale=0.49]{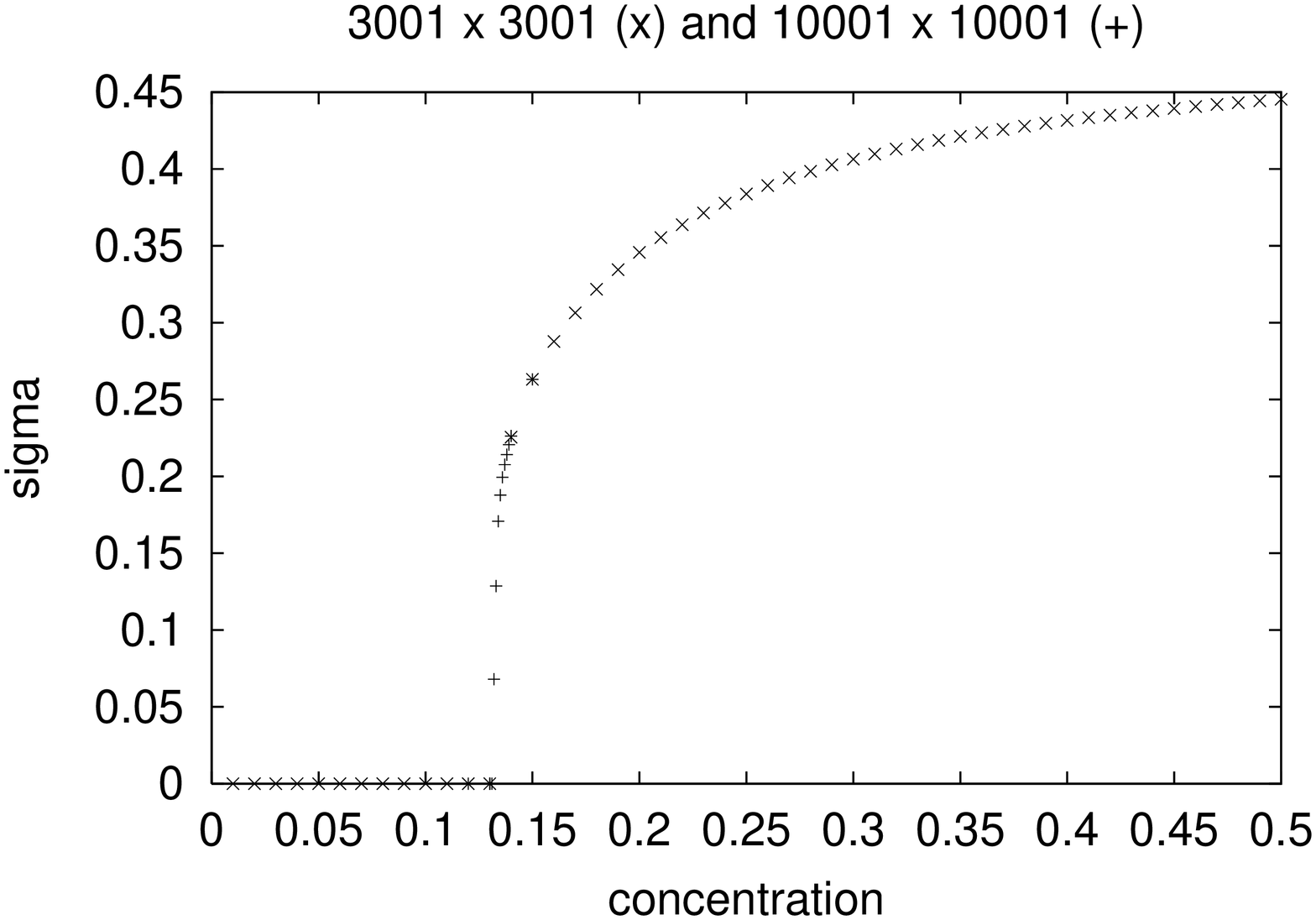}
\includegraphics[angle=-90,scale=0.49]{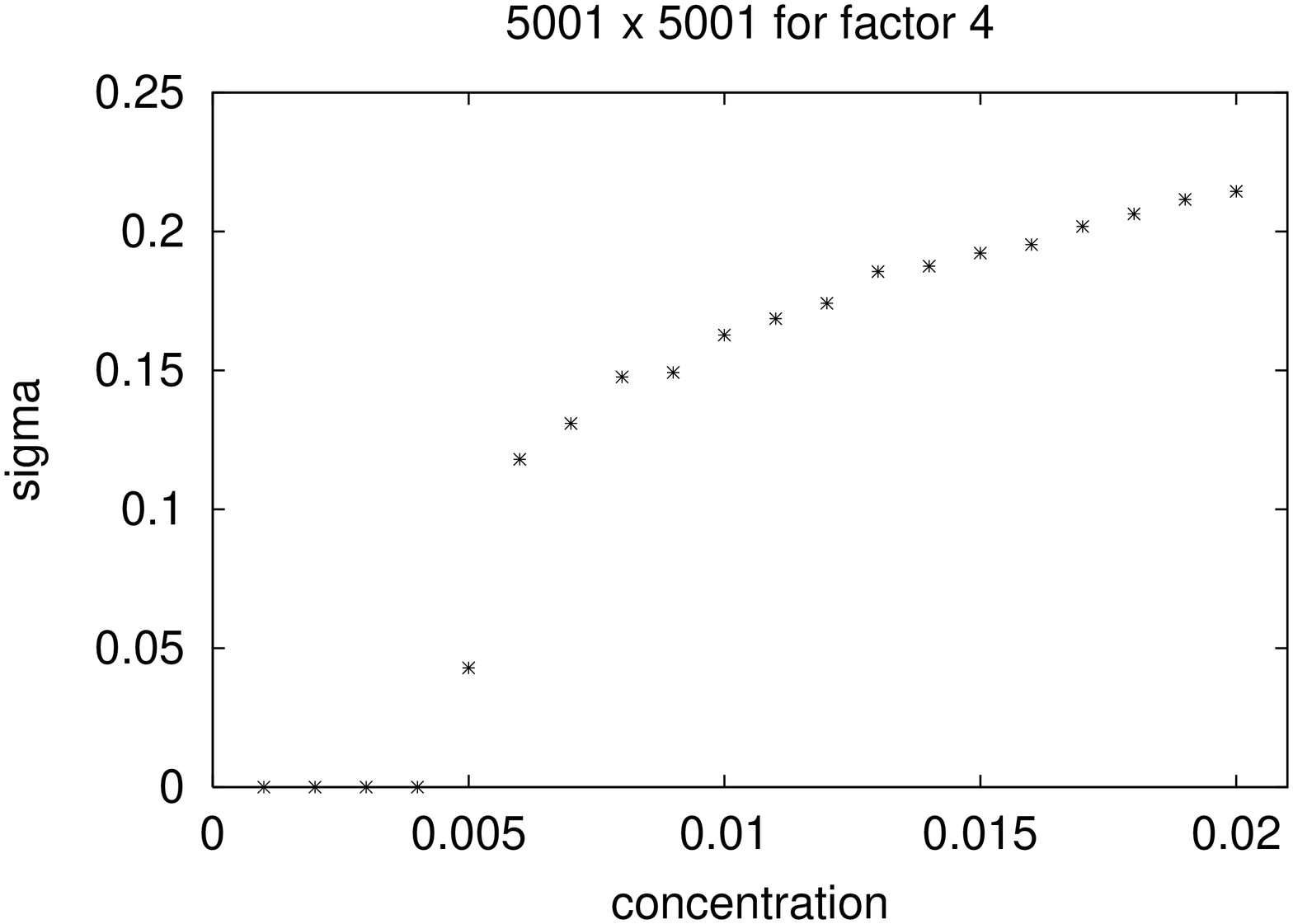}
\end{center}
\caption{Order parameter $\sigma$ versus population density $p$ for factor
2 (part a) and factor 4 (part b).
}
\end{figure}

\begin{figure}[hbt]
\begin{center}
\includegraphics[angle=-90,scale=0.4]{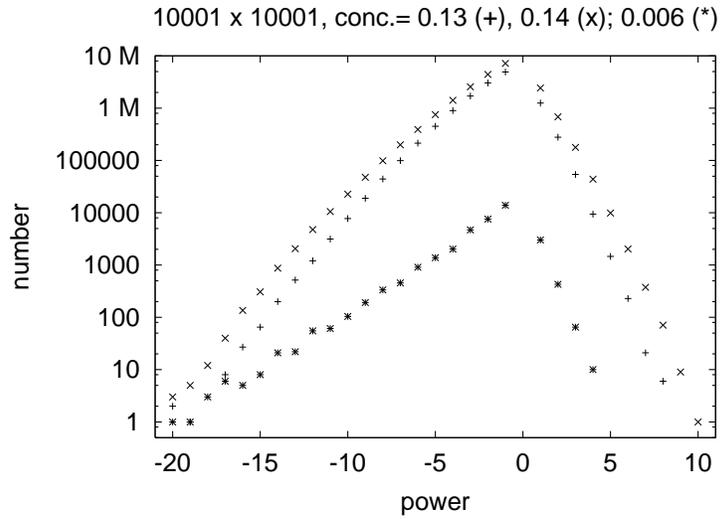}
\end{center}
\caption{Histogram of people having a certain power, as measured by $20h/H$.
The asymmetry factor is 2 for $p = 0.13$ below the transition (+) and for
$p = 0.14$ above the transition (x), while it is 4 for $p = 0.006$ near the 
transition (stars).
}
\end{figure}

\begin{figure}[hbt]
\begin{center}
\includegraphics[angle=-90,scale=0.4]{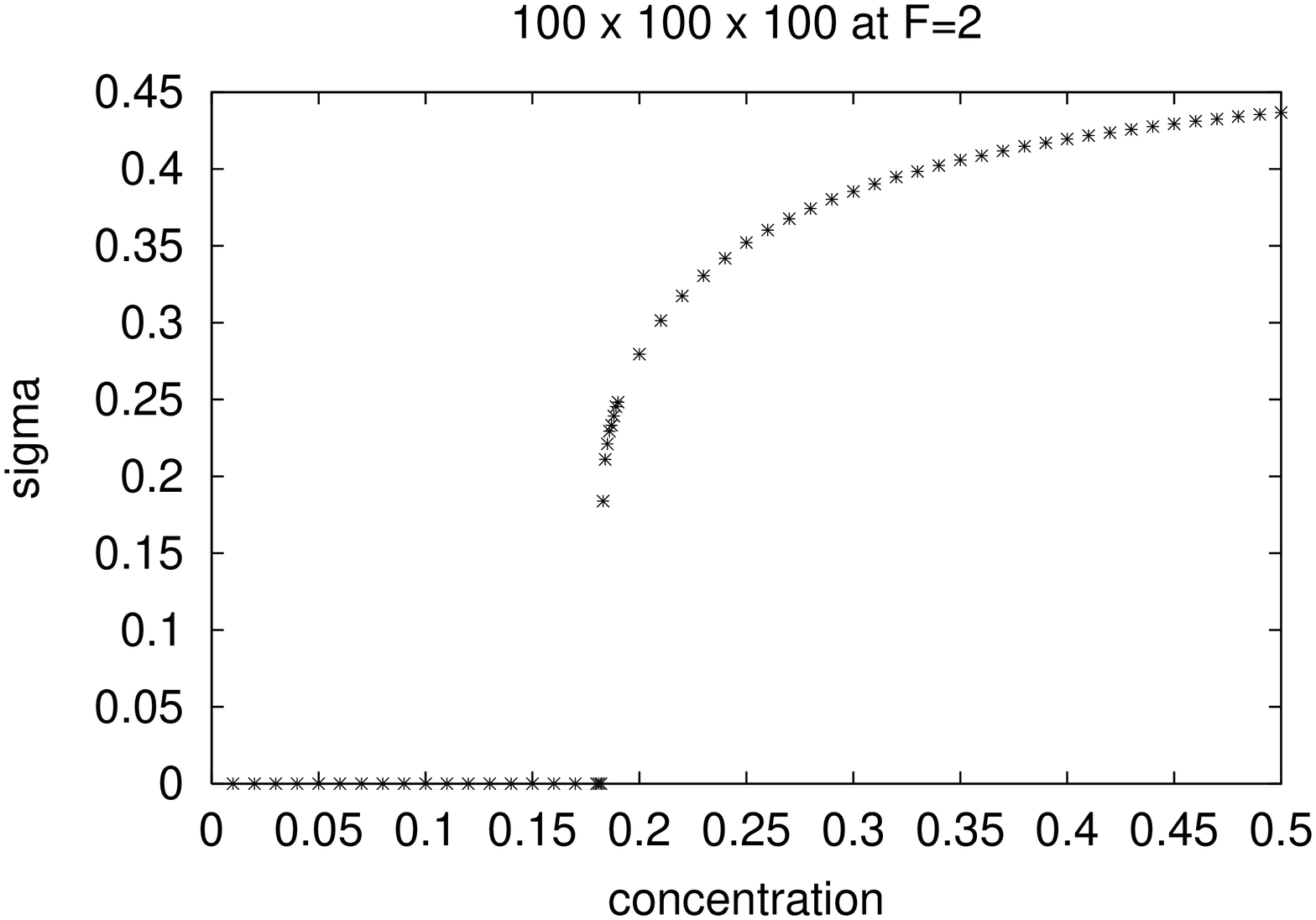}
\end{center}
\caption{As Fig.1a but in three dimensions.
}
\end{figure}

In the previous version of the Bonabeau model \cite{stauffer}, people diffuse
randomly on a big square lattice, on which they have a constant density 
$0 < p < 1$. If one agent $i$ wants to move onto a site already occupied by
another agent $k$, then a fight takes place. It is won with probability $q$
(see below) by agent $i$, and with probability $1-q$ by agent $k$. If $k$
wins, both stay at their place; if $k$ wins the two people exchange their
positions. The standard deviation $\sigma$ of the probabilities is the
order parameter: 
$$ \sigma^2 = \; <q^2> \; - \; <q>^2  \quad .\eqno(1)$$ 
Each agent $i$ keeps in mind a history $h(i)$ which is increased by one for 
every victory and decreased by one for every loss; moreover, the current value 
of $h$ is diminished by ten percent after every time step. One time step means 
that on average every agent is selected randomly once to move and possibly
fight. Initially $\sigma$ and all $h$ are zero. 
The above probability $q$ for $i$ to win against $k$ is 
 $$q = 1/(1 + \exp(\sigma[h(k)-h(i)]) \quad .\eqno (2)$$ 
Thus a global feedback exist through the order parameter $\sigma$ entering Eq.
(2). 

This model is by definition completely symmetric with respect to victories 
and losses. We measure the power of each agent on a scale from --20 to +20
through the ratio $20h/H$ where $H$
is the maximal value of $|h(i)|$ at that moment
in the whole population. Thus we have on average as many people of power --10
as with power +10. This unrealistic feature is now avoided by decreasing $h$ 
after a loss by an amount
$F > 1$ while increasing it after a victory only by 1. This
factor $F$ thus gives more weight to a loss than to a victory, just as in some
football tournaments a loss means the end of participation, while a victory 
only allows to proceed to the next round.

This asymmetry decreases the phase transition in the concentration $p$ from
0.32 for the old case $F=1$ to 0.134 at $F=2$ and 0.006 at
$F=4$, Fig.1. And it produces an asymmetry visible in Fig.2 which already
for a factor $F=2$ gives six times less people with positive power
than with negative power. The few people with strongly negative power might
be jail inmates. We made 200 or 500 iterations to get equilibrium. Note that
the power distribution hardly notices the phase transition, as seen by comparing
+ with x in Fig. 2. Fig.3 shows that also on the simple cubic lattice a phase
transition exists; the power distribution then also looks similar to Fig.2
(not shown).

\medskip
\noindent {\bf Acknowledgements}: We thank S. Moss de Oliveira 
for discussion,  PMMH at ESPCI for the warm
hospitality and Sorin T\u{a}nase-Nicola for help with the computer facilities. 
JSSM acknowledges funding from the Brazilian agencies CNPq and FAPERJ.

\end{document}